\documentclass{article}
\usepackage[T1]{fontenc}
\usepackage{geometry}
\usepackage{graphics}

\makeatletter

\newcommand{\LyX}{L\kern-.1667em\lower.25em\hbox{Y}\kern-.125emX\@}

\def\be{\begin{equation}}
\def\ee{\end{equation}}

\def\epi{\end{picture}}
\def\emi{\end{minipage}}

\def\QCD{{\rm QCD}}
\def\Born{{\rm Born}}

\def\inel{{\rm inel}}

\makeatother

\begin{document}

\title{\textbf{A New Approach to Nuclear Collisions at RHIC Energies}}

\author{H.J. DRESCHER\protect\( ^{1}\protect \), M. HLADIK\protect\( ^{1,3}\protect \),
S. OSTAPCHENKO\protect\( ^{2,1}\protect \), K. WERNER\protect\( ^{1}\protect \)\\
\\
 \textit{\protect\( ^{1}\protect \) SUBATECH, Université de Nantes -- IN2P3/CNRS
-- EMN,  Nantes, France }\\
\textit{\protect\( ^{2}\protect \) Moscow State University, Institute of Nuclear
Physics, Moscow, Russia}\\
\textit{\protect\( ^{3}\protect \) now at SAP AG, Berlin, Germany}}

\maketitle
\begin{abstract}
We present a new parton model approach for nuclear collisions at RHIC energies
(and beyond). It is a selfconsistent treatment using the same formalism for
calculating cross sections like \( \sigma _{\mathrm{tot}} \) or \( \sigma _{\mathrm{inel}} \)
and, on the other hand, particle production. Actually, the latter one is based
on an expression for the total cross section, expanded in terms of cut Feynman
diagrams. Dominant diagrams are assumed to be composed of parton ladders between
any pair of nucleons, with ordered virtualities from both ends of the ladder.
\end{abstract}

\section{Introduction}

The standard parton model approach to proton-proton or also nucleaus-nucleus
scattering amounts to presenting the partons of projectile and target by momentum
distribution functions, \( f_{A} \) and \( f_{B} \), and calculating inclusive
cross sections as \cite{sjo87, dur87}

\[
\sigma _{\mathrm{incl}}=\sum _{ij}\int dt\int dx^{+}\int dx^{-}f^{i}_{A}(x^{+},Q^{2})f_{B}^{j}(x^{-},Q^{2})\frac{d\hat{\sigma }_{ij}}{dt}(x^{+}x^{-}s),\]
where \( d\hat{\sigma }_{ij}/dt \) is the elementary parton-parton cross section.
This simple factorisation formula is the result of cancellations of complicated
diagrams (AGK cancellations) and hides therefore the complicated multiple scattering
structure of the reaction. The most obvious manifestion of such a structure
is the fact that the inclusive cross section exceeds the total one, so the average
number of elementary interactions must be bigger than one. The usual solution
is the so-called eikonalization, which amounts to re-introducing multiple scattering,
however, based on the above formula for the inclusive cross section. The latter
formula has the disdvantage of having lost many important details about the
multiple scattering aspects. Constructing in this way a model (in particular
an event generator) for particle production is completely arbitrary. For example,
energy obviously must be conserved, but the QCD formulas will not provide the
slightest hint how to realize energy conservation in a particular multiple scattering
event.

This problem has first been discussed in \cite{abr92},\cite{bra90}. The authors
claim that following from the nonplanar structure  of the corresponding diagrams,
conserving energy and momentum in a consistent way is crucial, and therefore
the incident energy has to be shared between the different elementary interactions,
both real and virtual ones. 

Following these ideas, we provide in this paper a rigorous treatment of the
multiple scattering aspect, such that questions as energy conservation are clearly
determined by the rules of field theory, removing the arbitrariness of the procedures
applied so far. The general idea is as follows: starting point is an expression
for the inelastic cross section for nucleus-nucleus scattering at very high
energies, expressed in terms of cut Feynman diagrams. Clearly at this point
certain assumptions concerning the dominace of certain classes of diagrams is
employed. But this assumptions really define the model, what follows is just
the application of the rules of field theory, calculating diagrams, making partial
summations, and finally interpreting these partial sums as ``partial cross
sections'' for certain physical processes. Production of particles (on the
parton level) is thus completely determined \cite{wer97}.

\section{The Elementary Cut Diagram: the Parton Ladder}

We want to write down an expression for the inelasic cross secion in nucleus-nucleus
(including nucleon-nucleon) scattering in terms of cut Feynman diagrams. We
will assume that dominant contributions will come from certain classes of diagrams,
which are composed of so-called ``elementary diagrams'', also referred to
as ``elementary interactions''. We will therefore first dicuss the elementary
diagrams, before introducing the multiple scattering theory. A complete diagram
will certainly contain cut and uncut elementary diagrams. We will investigate
explicitely only cut diagrams and use a relation between cut and uncut diagrams
to calculate the latter ones.

\begin{figure}[h]
{\par\centering \resizebox*{!}{0.3\textheight}{\includegraphics{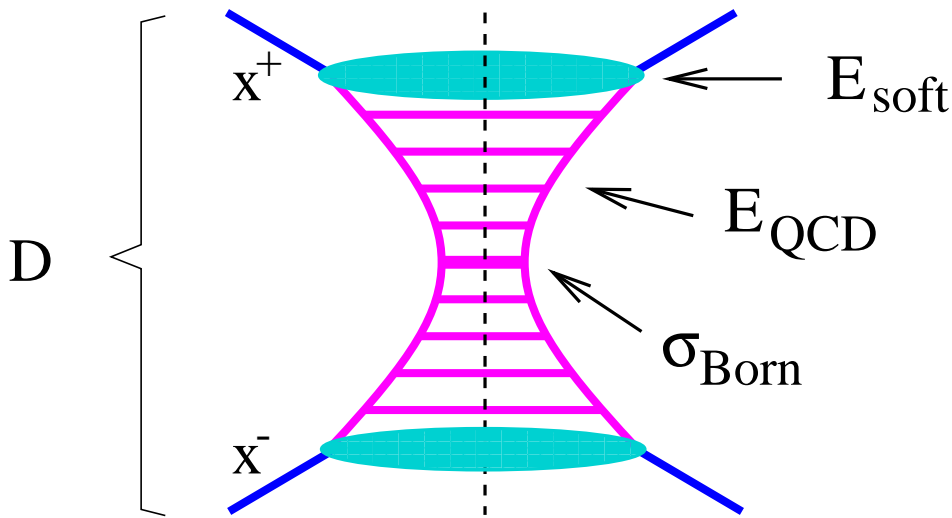}} \par}

\caption{The elementary cut diagram: parton ladder plus ``soft ends''.\label{d}}
\end{figure}

We assume an elementary interaction to be represented by a parton ladder with
``soft ends'', see fig \ref{d}.
The central part is a parton ladder with ordered virtualities, such that the
highest virtuality is at the center and the virtualities are decreasing towards
the ends of the ladder. This part of the diagram can be calculated using perturbative
techniques of QCD. Since the virtualities are decreasing towards the ends, one
reaches finally values where perturbative calculation can no longer be employed,
although the longitudinal momentum fraction of the corresponding parton may
be much smaller than one. This means there is still a large mass ``object''
between the first parton of the ladder and the nucleon, however, with small
virtualities involved \cite{lan94}. The most naturel candidate for such an
object is the soft Pomeron, which can not be calculated from first principles,
but where reasonable parametrisations exist, based on general considerations
of scattering matrices in the limit of very high energies. So, the mathematical
expression corresponding to the cut diagram of fig. \ref{d} is
\begin{equation}
\label{dsemi}
D_{\mathrm{semi}}(s,x^{+},x^{-},b)=\int db'\sum _{ij}\int \frac{dx^{+}_{1}}{x^{+}_{1}}\int \frac{dx^{-}_{1}}{x^{-}_{1}}E_{\mathrm{soft}}^{\mathrm{i}}\left( \frac{x^{+}_{1}}{x^{+}},b'\right) E_{\mathrm{soft}}^{\mathrm{j}}\left( \frac{x^{-}_{1}}{x^{-}},b-b'\right) \sigma _{\mathrm{ladder}}^{ij}(x^{+}_{1}x^{-}_{1}s),
\end{equation}
where \( E_{\mathrm{soft}} \) represents the soft Pomeron at each end of the
ladder and \( \sigma _{\mathrm{ladder}}^{ij} \) the parton ladder itself, the
precise definition of both quantities being given in the following. 

The hard part of the elementary interaction, \( \sigma _{\mathrm{ladder}}^{ij} \),
is given as
\begin{eqnarray}
\sigma _{\mathrm{ladder}}^{ij}(\hat{s}) & = & \sum _{kl}\int dw^{+}dw^{-}dQ_{1}^{2}\\
 &  & E_{\mathrm{QCD}}^{ik}(Q_{0}^{2},Q_{1}^{2},w^{+})\, E_{\mathrm{QCD}}^{jl}(Q_{0}^{2},Q_{1}^{2},w^{-})\, {d\sigma _{\mathrm{Born}}^{kl}\over dQ^{2}}(w^{+}w^{-}\hat{s},Q_{1}^{2}),\nonumber 
\end{eqnarray}
 where \( d\sigma _{\mathrm{Born}}^{kl}/dQ^{2} \)represents the hardest scattering
in the middle of the ladder (indicated symbolically by the somewhat thicker
ladder rung in fig. \ref{d}), 
\begin{figure}[h]
{\par\centering \resizebox*{0.7\textwidth}{!}{\includegraphics{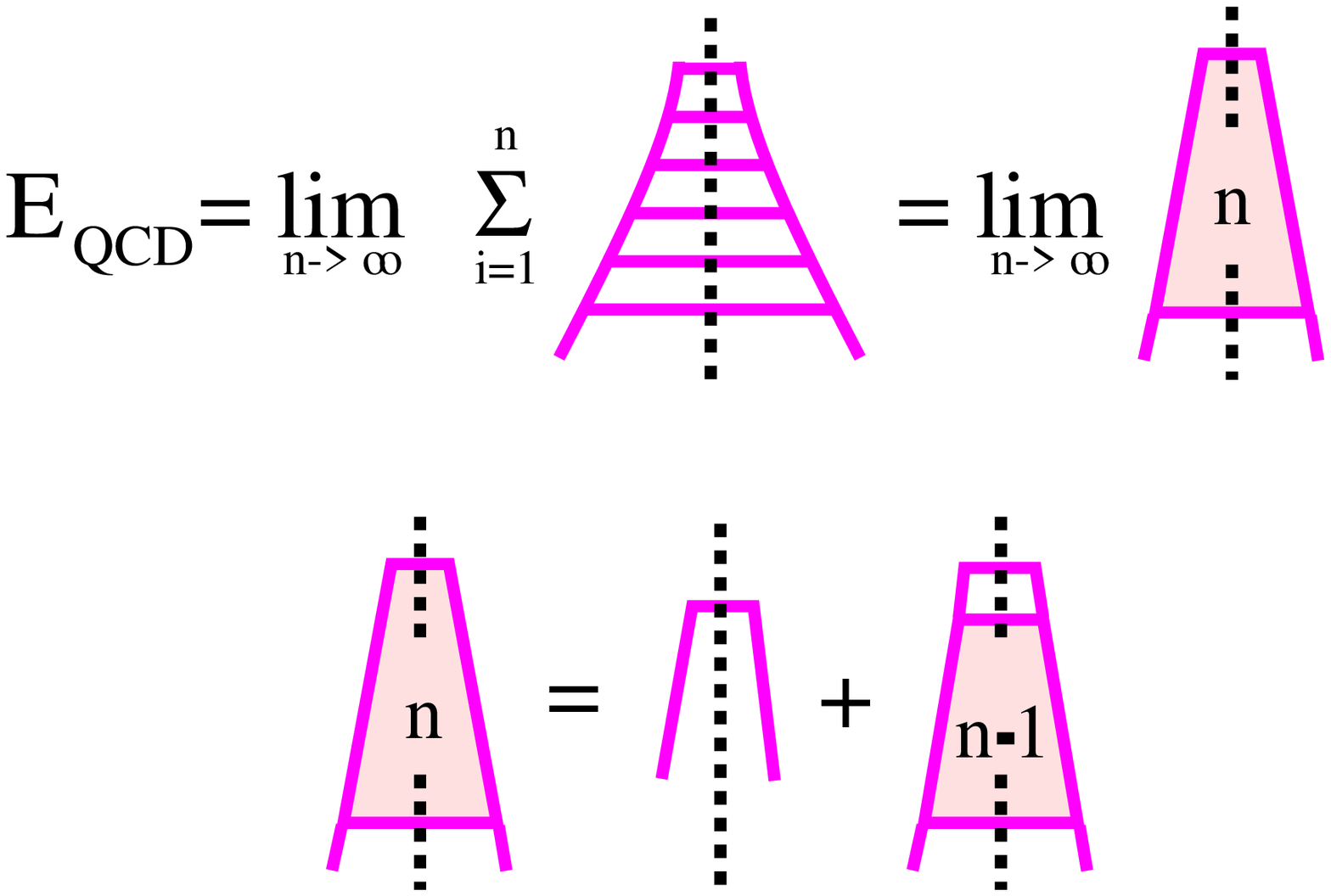}} \par}

\caption{The calculation of \protect\( E_{\mathrm{QCD}}\protect \).\label{evol}}
\end{figure}
and where \( E_{\mathrm{QCD}} \) represents the evolution of parton cascade
from scale \( Q_{0}^{2} \) to \( Q_{1}^{2} \), using the DGLAP approximation
\cite{alt82, ell96}, given as (see fig. \ref{evol})

\begin{equation}
E^{ij}_{\QCD }(Q_{0}^{2},Q_{1}^{2},x)=\lim _{n\to \infty }E^{(n)ij}_{\QCD }(Q_{0}^{2},Q_{1}^{2},x),
\end{equation}
 where \( E_{\mathrm{QCD}}^{(n)} \) represents an ordered ladder with at most
\( n \) ladder rungs. This is  calculated iteratively based on  
\begin{eqnarray}
 &  & E_{\QCD }^{(n)ij}(Q_{0}^{2},Q_{1}^{2},x)=\delta (1-x)\, \delta _{ij}\, \Delta ^{i}(Q_{0}^{2},Q_{1}^{2})\label{eqcd} \\
 &  & +\sum _{k}\int _{Q_{0}^{2}}^{Q_{1}^{2}}\frac{dQ^{2}}{Q^{2}}\int _{0}^{1-\epsilon }{d\xi \over \xi }\, {\alpha _{s}\over 2\pi }\, E_{\QCD }^{(n-1)ik}(Q_{0}^{2},Q^{2},\xi )\, \Delta ^{j}(Q^{2},Q_{1}^{2})\, P_{k}^{j}({x\over \xi }),\nonumber \label{eqcd} 
\end{eqnarray}
 where the indices \( i \), \( j \), \( k \) represent parton flavors. \( P_{k}^{j} \)
are the Altarelli-Parisi splitting functions and \( \Delta ^{j} \) is the so-called
Sudakov form factor. The soft part of the elementary interaction, \( E_{\mathrm{soft}} \),
is the usual soft Pomeron expression.

In addition to the semihard contribution \( D_{\mathrm{semi}} \), one has to
consider the expression representing the purely soft contribution:
\begin{equation}
\label{softampl}
D_{\mathrm{soft}}\left( s,x^{+},x^{-},b\right) =E_{\mathrm{soft}}\! \left( \frac{s_{0}}{x^{+}x^{-}s},b\right) ,
\end{equation}
with the scale parameter \( s_{0}=1 \) GeV. The complete contribution, representing
an elementary  inelastic interaction in an energy range of say \( 10 \) - \( 10^{4} \)
GeV, is therefore  given as 
\begin{equation}
\label{gtot}
D=D_{\mathrm{soft}}+D_{\mathrm{semi}}.
\end{equation}
 We would like to stress, that the ``soft end'' of the semihard Pomeron has
 exactly the same structure as the soft contribution itself,  no new parameters
enter. 

There is still something missing: the outer legs of the elementary diagram are
not the nucleons, but nucleon ``constituents'', to be more precise quark-antiquark
pairs. We call these constituents also ``participants'' to indicate that they
are actively participating in the interaction, in contrast to the remnants,
which represent the non-participating part of the nucleons. So for each incoming
leg, we have an additional factor \( F_{\mathrm{part}} \), which we assume
to be of the form \( F_{\mathrm{part}}(x)=x^{-\alpha _{\mathrm{part}}} \).
Similarly, for each remnant, we add a factor \( F_{\mathrm{remn}}(x)=x^{-\alpha _{\mathrm{remn}}} \),
where the arguments of \( F_{\mathrm{remn}}(x) \) are the momentum fractions
of the remnants. We define
\[
2G(s,x^{+},x^{-},b)=F_{\mathrm{part}}(x^{+})D(s,x^{+},x^{-},b)F_{\mathrm{part}}(x^{-}),\]
where we introduced a factor 2 for later convenience.

\section{Multiple Scattering}

\begin{figure}[h]
{\par\centering \resizebox*{1\textwidth}{!}{\includegraphics{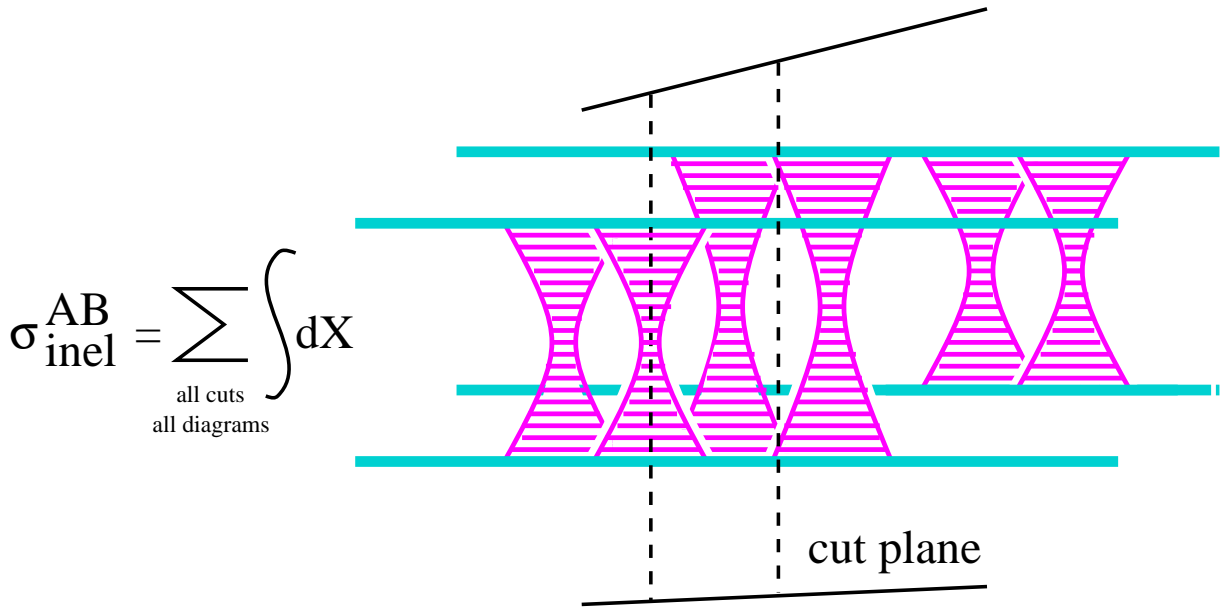}} \par}

\caption{The inelastic cross section for nucleus-nucleus scattering.\label{mult}}
\end{figure}

We assume that the dominant diagrams for nucleus-nucleus scattering are those
which consist of elementary diagrams as discussed in the previous sections,
see fig. \ref{mult}.
One easily writes down the corresponding formula: each cut Pomeron  contributes
a factor \( 2CG \), each uncut one a factor \( (-2CG) \)(the semihard Pomeron
amplitude is assumed to be imaginary), and each remnant \( F^{+} \) or \( F^{-} \).
So we get
\begin{eqnarray}
\sigma _{\mathrm{inel}}\! (s) & = & \int \! d^{2}b\int dT_{AB}\sum _{m_{1}l_{1}}\ldots \sum _{m_{AB}l_{AB}}\, \int \, \prod _{k=1}^{AB}\left\{ \prod ^{m_{k}}_{\mu =1}dx_{k,\mu }^{+}dx_{k,\mu }^{-}\, \prod ^{l_{k}}_{\lambda =1}d\tilde{x}_{k,\lambda }^{+}d\tilde{x}_{k,\lambda }^{-}\right\} \nonumber \\
 &  & \prod _{k=1}^{AB}\left\{ {1\over m_{k}!}{1\over l_{k}!}\, \prod _{\mu =1}^{m_{k}}2CG(s,x_{k,\mu }^{+},x_{k,\mu }^{-},b)\, \prod _{\lambda =1}^{l_{k}}-2CG(s,\tilde{x}_{k,\lambda }^{+},\tilde{x}_{k,\lambda }^{-},b)\right\} \nonumber \\
 &  & \prod _{i=1}^{A}F^{+}\! \left( x_{i}^{P+}-\sum _{\pi (k)=i}\tilde{x}_{k,\lambda }^{+}\right) \, \prod _{j=1}^{B}F^{-}\! \left( x_{j}^{T-}-\sum _{\tau (k)=j}\tilde{x}_{k,\lambda }^{-}\right) \label{s1} 
\end{eqnarray}
with
\[
x_{i}^{P+}=1-\sum _{\pi (k)=i}x_{k,\mu }^{+}\]
\[
x_{i}^{T-}=1-\sum _{\tau (k)=j}x_{k,\mu }^{-},\]
where \( \int dT_{AB} \) represents the integration over transverse coordinates
of projectile and target nucleons with the appropriate weight given by the so-called
thickness functions \cite{wer93}. The factor \( D_{AB} \) is given as 
\[
D_{AB}=\prod _{i=1}^{A}\left\{ \frac{1}{\sqrt{C}}+(1-\frac{1}{\sqrt{C}}\delta ^{+}_{i})\right\} \prod _{j=1}^{B}\left\{ \frac{1}{\sqrt{C}}+(1-\frac{1}{\sqrt{C}}\delta ^{-}_{j})\right\} ,\]
with
\[
\delta ^{\pm }_{n}=\left\{ \begin{array}{ll}
1 & \mathrm{if}\, \mathrm{nucleon}\, \mathrm{is}\, \mathrm{passive}\\
0 & \mathrm{if}\, \mathrm{nucleon}\, \mathrm{is}\, \mathrm{active}
\end{array}\right. ,\]
where an active nucleon participates in at least one elementary interaction,
whereas a passive one does not. The functions \( \pi (k) \) and \( \tau (k) \)
refer to the projectile and the target nucleons participating in the \( k^{\mathrm{th}} \)
interaction. We fully account for energy-momentum conservation, which we consider
 extremely important due to the nonplanarity of the diagrams, implying the interactions
to occur in parallel. 

The expansion of \( \sigma _{\mathrm{inel}} \) in terms of cut diagrams as
given in eq. \ref{s1} represents a sum of a large number of positive and negative
terms, including all kinds of interferences, which excludes any probabilistic
interpretation. Our strategy consists therefore of performing partial summations
such that the remaining terms allow such an interpretation \cite{agk73}. So
we classify the diagrams according to the cut elementary diagrams (real emissions),
and then sum over all diagrams of a given class, which amounts to summing over
uncut elementary diagrams (sum over virtual emissions):
\begin{eqnarray*}
\sigma _{\mathrm{inel}} & = & \sum _{\mathrm{cut}\, \mathrm{diagrams}\, \mathcal{D}}\int dX\, d\tilde{X}\, \mathcal{D}\\
 & = & \sum _{\mathrm{cut}\, \mathrm{ladders}}dX\left\{ \sum _{\mathrm{uncut}\, \mathrm{ladders}}\int d\tilde{X}\, \mathcal{D}\right\} ,
\end{eqnarray*}
where \( \mathcal{D} \) is the mathematical expression corresponding to a diagram
as shown in fig. \ref{mult}, appearing in formula \ref{s1}, and where \( X \)
and \( \tilde{X} \) represents all the light cone momenta of the cut and uncut
elementary diagrams. The term in brackets \( \left\{ \ldots \right\}  \) may
be finally interpreted as probability for the corresponding ``ladder configuration''. 

Let us write the formulas explicitely. We have  
\begin{eqnarray}
\sigma _{\mathrm{inel}}\! (s) & = & \int \! d^{2}b\int dT_{AB}\sum _{m_{1}}\ldots \sum _{m_{AB}}\, \int \, \prod _{k=1}^{AB}\left\{ \prod ^{m_{k}}_{\mu =1}dx_{k,\mu }^{+}dx_{k,\mu }^{-}\right\} \nonumber \\
 &  & \prod _{k=1}^{AB}\left\{ \frac{1}{m_{k}!}\, \prod _{\mu =1}^{m_{k}}2G(s,x_{k,\mu }^{+},x_{k,\mu }^{-},b)\right\} \; R\left( s,x^{P+},x^{T-},b\right) ,\label{sigmanucl} 
\end{eqnarray}
  with 
\begin{eqnarray}
R\left( s,x^{P+},x^{T-},b\right)  & = & \sum _{l_{1}}\ldots \sum _{l_{AB}}\, \int \, \prod _{k=1}^{AB}\left\{ \prod ^{l_{k}}_{\lambda =1}d\tilde{x}_{k,\lambda }^{+}d\tilde{x}_{k,\lambda }^{-}\right\} \; \prod _{k=1}^{AB}\left\{ \frac{1}{l_{k}!}\, \prod _{\lambda =1}^{l_{k}}-2G(s,\tilde{x}_{k,\lambda }^{+},\tilde{x}_{k,\lambda }^{-},b)\right\} \nonumber \label{r} \\
 & \times  & \prod _{i=1}^{A}F^{+}\! \left( x_{i}^{P+}-\sum _{\pi (k)=i}\tilde{x}_{k,\lambda }^{+}\right) \prod _{j=1}^{B}F^{-}\left( x^{T-}_{j}-\sum _{\tau (k)=j}\tilde{x}_{k,\lambda }^{-}\right) .\label{rremnant} 
\end{eqnarray}
 The variables appearing in eq. (\ref{sigmanucl}) may be represented by two
multivariables: the interaction-type variable \( M \) which specifies for each
of the \( AB \) nucleon pairs the type of the interaction (how many cut Pomerons
of which type occur), and the momentum variable \( X \), already mentoned earlier,
which specifies for each elementary interaction the momentum fractions. Eq.
(\ref{sigmanucl}) may thus be written as
\begin{equation}
\label{siginelab3}
\sigma _{\mathrm{inel}\textrm{ }}=\sum _{M}\, \int dX\, \Omega (M,X).
\end{equation}
 where \( \Omega (M,X) \) is the integrand of eq. (\ref{sigmanucl}). Both
variables, \( K=\{M,X\} \) represent a ``ladder configuration'' and \( \Omega (M,X) \)
considered to be the corresponding probability density.

There are two fundamental problems to be solved:

\begin{itemize}
\item the sum over virtual emmisions has to be performed
\item tools have to be delevopped to deal with the multidimensional probability distribution
\( \Omega (M,X) \). 
\end{itemize}
Both are difficult tasks. There is no way to do the summation numericly in case
of two heavy nuclei \( A \) and \( B \): we have \( AB \) summation indices,
so if we assume 10 terms per index to reach convergence, we have to sum over
\( 10^{AB} \) terms! It is also out of question to use Monte Carlo methods,
since positive and negative terms occur. However, we are able to provide a solution,
as discussed later. Concerning the multidimensional probability distribution
\( \Omega (M,X) \), we are going to develop methods well known in statistical
physics (Markov chain techniques), which we also are going to discuss in detail
later. So finally, we are able to calculate the probability distribution \( \Omega (M,X) \),
and are able to generate (in a Monte Carlo fashion) ``ladder configurations''
\( (M,K) \) according to this probability distribution. The next task amounts
to generate explicitely partons, again based on our master formula eq. \ref{sigmanucl}.
This will be discussed in next section.

\section{Parton Configurations}

In this section, we consider the generation of parton configurations in nucleus-nucleus
(including proton-proton) scattering for a given ladder configuration, which
means, the number of elementary interactions per nucleon-nucleon pair is known,
as well as the light cone momentum fractions \( x^{+} \) and \( x^{-} \) of
each elementary interaction. A parton configuration is specified by the number
of partons, their types and momenta. We showed earlier that the inelastic cross
section may be written as

\begin{equation}
\sigma _{\inel }=\sum _{K\in \cal K}\Omega (K),
\end{equation}
 where the symbol \( \Sigma  \) means \( \Sigma \int  \) and where \( K=\{M,X\} \)
represents a ladder configuration. The function \( \Omega (K) \) is known (see
eq. (\ref{sigmanucl})) and is interpreted as probability distribution for a
ladder configuration \( K \). For each individual ladder a term \( 2CG \)
appears in the formula for \( \Omega (K) \), where \( 2G \) itself can be
expressed in terms of parton configurations, which provides probability distributions
for parton configurations, and which provides the basis for generating partons.
We want to stress that the parton generation is also based on the master formula
eq. (\ref{sigmanucl}), no new elements enter. In the following, we want to
discuss in detail the generation of parton configurations for an elementary
interaction with given light cone momentum fractions \( x^{+} \) and \( x^{-} \)and
given impact parameter difference \( b \) between the corresponding pair of
interacting nucleons. 

First, we have to specify the type of elementary interaction (soft or semihard).
The corresponding probabilities are

\begin{equation}
G_{\mathrm{semi}}(s,x^{+},x^{-},b)\, /\, G(s,x^{+},x^{-},b)
\end{equation}
 and
\begin{equation}
G_{\mathrm{soft}}(s,x^{+},x^{-},b)\, /\, G(s,x^{+},x^{-},b)
\end{equation}
 respectively.

Let us now consider a semihard contribution. We obtain the desired probability
distributions from the explicit expressions for \( 2G_{\mathrm{semi}} \). For
given \( x^{+} \), \( x^{-} \), we have
\begin{equation}
\label{2cg}
2G_{\mathrm{semi}}(s,x^{+},x^{-},b)\propto \int dx_{1}^{+}dx_{1}^{-}\left\{ \int \! d^{2}b'\sum _{ij}\, E^{i}_{\mathrm{soft}}(\frac{x_{1}^{+}}{x^{+}},b')\, E^{j}_{\mathrm{soft}}(\frac{x_{1}^{-}}{x^{-}},b-b')\, \sigma _{\mathrm{ladder}}^{ij}(x_{1}^{+}x_{1}^{-}s)\right\} ,
\end{equation}
 with 
\begin{eqnarray}
\sigma _{\mathrm{ladder}}^{ij}(\hat{s}) & = & \sum _{kl}\int dw^{+}dw^{-}dQ^{2}\\
 &  & E_{\mathrm{QCD}}^{ik}(Q_{0}^{2},Q^{2},w^{+})\, E_{\mathrm{QCD}}^{jl}(Q_{0}^{2},Q^{2},w^{-})\, {d\sigma _{\mathrm{Born}}^{kl}\over dQ^{2}}(w^{+}w^{-}\hat{s},Q^{2}),\nonumber 
\end{eqnarray}
 representing the perturbative parton-parton cross section, where both initial
partons are taken at the virtuality \( Q^{2}_{0} \). The integrand \( \left\{ \ldots \right\}  \)
of eq. \ref{2cg} serves as probability distribution to generate \( x_{1}^{+} \)
and \( x_{1}^{-} \). 

Knowing the momentum fractions \( x_{1}^{+} \) and \( x_{1}^{-} \)of the ``first
partons'' of the parton ladder, we can construct the complete ladder. To do
so, we generalize the definition of the parton-parton cross section \( \sigma _{\mathrm{ladder}} \)
to arbitrary virtualities of the initial partons, we define 
\begin{eqnarray}
\sigma _{\mathrm{ladder}}^{ij}(Q_{1}^{2},Q_{2}^{2},\hat{s}) & = & \sum _{kl}\int dw^{+}dw^{-}dQ^{2}\\
 &  & E_{\QCD }^{ik}(Q_{1}^{2},Q^{2},w^{+})\, E_{\QCD }^{jl}(Q_{2}^{2},Q^{2},w^{-})\, {d\sigma _{\Born }^{kl}\over dQ^{2}}(w^{+}w^{-}\hat{s},Q^{2}).\nonumber 
\end{eqnarray}
 and 
\begin{eqnarray}
\sigma _{\textrm{ord}}^{ij}(Q_{1}^{2},Q_{2}^{2},\hat{s}) & = & \sum _{k}\int dw^{-}dQ^{2}\\
 &  & E_{\QCD }^{jk}(Q_{2}^{2},Q^{2},w^{-})\, \Delta ^{i}(Q_{1}^{2},Q^{2})\, {d\sigma _{\Born }^{ki}\over dQ^{2}}(w^{-}\hat{s},Q^{2}).\nonumber 
\end{eqnarray}
 representing ladders with ordering of virtualities on both sides  (\( \sigma _{\textrm{jet}} \))
or on one  side only (\( \sigma _{\textrm{ord}} \)).  We calculate and tabulate
\( \sigma _{\textrm{jet}} \) and \( \sigma _{\textrm{ord}} \) initially so
that we can use them via interpolation to generate partons. The generation of
partons is done in an iterative fashion based on  the following equations:  
\begin{eqnarray}
\sigma _{\mathrm{ladder}}^{ij}(Q_{1}^{2},Q_{2}^{2},\hat{s}) & = & \sum _{k}\int \frac{dQ^{2}}{Q^{2}}\int \frac{d\xi }{\xi }\, \Delta ^{i}(Q_{1}^{2},Q^{2})\, \frac{\alpha }{2\pi }\, P_{i}^{k}(\xi )\, \sigma _{\mathrm{ladder}}^{kj}(Q^{2},Q_{2}^{2},\xi \hat{s})\\
 &  & +\sigma _{\mathrm{ord}}^{ij}(Q_{1}^{2},Q_{2}^{2},\hat{s}).\nonumber 
\end{eqnarray}
and
\begin{eqnarray}
\sigma _{\mathrm{ord}}^{ij}(Q_{1}^{2},Q_{2}^{2},\hat{s}) & = & \sigma _{\mathrm{Born}}^{ij}(Q_{1}^{2},Q_{2}^{2},\hat{s})\\
 &  & +\sum _{k}\int \frac{dQ^{2}}{Q^{2}}\int \frac{d\xi }{\xi }\, \Delta ^{i}(Q_{1}^{2},Q^{2})\, \frac{\alpha }{2\pi }\, P_{i}^{k}(\xi )\, \sigma _{\mathrm{ladder}}^{kj}(Q^{2},Q_{2}^{2},\xi \hat{s}).\nonumber 
\end{eqnarray}

\section{Outlook}

So far we presented a consistent and very transparent new approch to calculate
parton production in nucleus-nucleus (including nucleon-nucleon) scattering.
But, unfortunately, the real world consists of hadrons, so we still have to
deal with the problem of hadronization. This is not so clear. We provide a ``minimal
model'' where we simply translate the partons from each individual elementary
interaction into the language of relativistic strings, the latter ones being
decayed using the machinery of relativistic string decay. An alternative would
be to take our partons as initial condition for a transport treatment of a partonic
system. We do not want to explore these options any further in this paper.

\section{Acknowledgements}

This work has been funded in part by the IN2P3/CNRS (PICS 580) and the Russian
Foundation of Basic Researches (RFBR-98-02-22024).

\end{document}